\RequirePackage{tikz}

\documentclass[]{sn-jnl}

\usepackage{tikz}
\usepackage{array}
\usepackage{float}
\usetikzlibrary{positioning}
\usetikzlibrary{shapes,arrows,arrows,positioning,fit}

\jyear{2021}%

\theoremstyle{thmstyleone}%
\theoremstyle{thmstyletwo}%

\theoremstyle{thmstylethree}%

\begin{document}

\title[Analyzing and Comparing Omicron Lineage Variants Protein-Protein Interaction Network using Centrality Measure ]{Analyzing and Comparing Omicron Lineage Variants Protein-Protein Interaction Network using Centrality Measure }

\author*[1]{\fnm{Mamata} \sur{Das}}\email{dasmamata.india@gmail.com}

\author*[1]{\fnm{Selvakumar} \sur{K}}\email{kselvakumar@nitt.edu}
\equalcont{These authors contributed equally to this work.}

\author[1]{\fnm{P.J.A.} \sur{Alphonse}}\email{alphonse@nitt.edu}
\equalcont{These authors contributed equally to this work.}

\affil[1]{\orgdiv{Department of Computer Applications}, \orgname{NIT Trichy}, \orgaddress{\street{NH 83}, \city{Trichy}, \postcode{620015}, \state{Tamil Nadu}, \country{India}}}

\abstract{The Worldwide spread of the Omicron lineage variants has now been confirmed. It is crucial to understand the process of cellular life and to discover new drugs need to identify the important proteins in a protein interaction network (PPIN). PPINs are often represented by graphs in bioinformatics, which describe cell processes. There are some proteins that have significant influences on these tissues, and which play a crucial role in regulating them. The discovery of new drugs is aided by the study of significant proteins. These significant proteins can be found by reducing the graph and using graph analysis. Studies examining protein interactions in the Omicron lineage (\textbf{B.1.1.529}) and its variants (\textbf{BA.5, BA.4, BA.3, BA.2, BA.1.1, BA.1}) are not yet available. Studying Omicron has been intended to find a significant protein. $68$ nodes represent $68$ proteins and $52$ edges represent the relationship among the protein in the network. A few centrality measures are computed namely page rank centrality (PRC), degree centrality (DC),closeness centrality (CC), and betweenness centrality (BC) together with node degree and Local Clustering Co-efficient(LCC). We also discover $18$ network clusters using Markov clustering. $8$ significant proteins (candidate gene of Omicron lineage variants) were detected among the $68$ proteins, including \textbf{AHSG, KCNK1, KCNQ1, MAPT, NR1H4, PSMC2, PTPN11} and, \textbf{UBE21} which scored the highest among the Omicron proteins. It is found that in the variant of Omicron protein-protein interaction networks, the \textbf{MAPT} protein's impact is the most significant.}


\maketitle

\section{Introduction}\label{sec1}
An impartial panel of scientists known as TAG-VE (Technical Advisory Group on SARS-CoV-2 Virus Evolution) regularly observes and examines the appraise of the SARS-CoV-2 virus to determine if specific mutations or combinations of mutations have an impact on the behaviour of the virus. The B.1.1.529 variants of SARS-CoV-2 was the subject of an evaluation by the TAG-VE on November 26, 2021. South Africa disclosed the B.1.1.529 version on November 24, 2021 to World Health Organization (WHO) \cite{10665-62429} for the first time. The WHO has classified B.1.1.529 as a VOC under the name Omicron considering the data that a negative shift in COVID-19 epidemiology has occurred. Similar to other SARS-CoV-2 variations, there are numerous lineages and sublineages in the Omicron variation. Omicron presently has $3$ main lineages: BA.5, BA.4, and BA.2. Omicron Pango Lineage currently has six different variations or sublineages (BA.5, BA.4, BA.3, BA.2, BA.1.1, BA.1). Although these lineages are frequently extremely similar to one another, there may be variations between lineages that influence how the virus behaves.
In our research, we have created seven PPI networks of Omicron Pango Lineage including all the variants. The network has been created on STRING, analyzed the network and find the most influential proteins from the network. The networks that describe the interactions between the parts of such complex systems are easier to analyse than it is to investigate each component separately. The placement of some significant or influential elements in most networks such as crucial proteins in PPI networks is a well-known fact in the analysis of biological and social networks. These locations, or vertices, have some unique structural characteristics. Such facts are quantified using various centrality metrics. The vertices and edges of a graph can be ranked from several perspectives based on centrality measurements. To pinpoint "central" nodes in extensive networks, numerous centrality measures (CM) have been developed. The user can choose whatever metric best fits the study of a certain network because there are several options available for ranking influential nodes. The effect of the network architecture on how influential nodes are ranked by centrality metrics further complicates the selection of an appropriate measure. In order to find the centrality metric that is most successful at predicting influential proteins, we looked at the centrality profiles of the nodes of Omicron PPINs. We looked at how a broad range of widely used centrality measurements reflects various topological network properties. This study demonstrates the state-of-the-art in biological network centrality estimations. In order to identify the most significant protein in the network, this research presents $4$ centrality metrics (page rank centrality (PRC), degree centrality (DC), closeness centrality (CC), and betweenness centrality (BC)) that are added with some significant scores (node degree and Local Clustering Co-efficient(CCo), and p-value) on Omicron variant's PPI networks.
\section{Related Work}\label{sec2}
Graph structures known as biological networks and social networks can be used to describe a variety of complex systems, including biological and social systems \cite{ghasemi2014centrality}. For determining significant functional characteristics of a network \cite{das2021markov}\cite{das2022analytical}, selecting an appropriate set of centrality measurements is essential. \cite{landherr2010critical} the paper has been considered in relation to a critical analysis of centrality measures in social networks. Three straightforward conditions for the behaviour of centrality measures were used to analyse certain centrality measures (BC, CC, DC, and eigenvector centrality). The author has been analysis of PPI using Skyline Query on Parkinson's disease \cite{diansyah2019analysis}. One of the disorders with the highest rate of global growth, Parkinson's disease, was shown to have 12 important proteins. The PPI network features have been represented by attributes based on centrality measures. The target genes for cancer illnesses were discovered by the author using protein-protein interaction networks \cite{amala2019identification}. Hubs and centrality measurements were used to examine the possible genes. They extract the genes with the highest scores in both mutation rates and graph centrality in order to identify the target genes. The author compared 27 popular centrality measurements using yeast PPINs \cite{ashtiani2018systematic}. The measurements classify and arrange the networks' influential nodes. They have also used hierarchical clustering and principal component analysis (PCA), and they discovered that the topology of the network affects which metrics are the most useful. The author has provided both historical and contemporary research on social network centrality measures in \cite{das2018study} survey paper. They discussed created centrality measurements and mathematical definitions. Additionally, they demonstrate various centrality measure uses in the fields of education research \cite{grunspan2014understanding}, biology \cite{ghasemi2014centrality}, traffic \cite{jayaweera2017centrality}, transportation \cite{wang2017novel}, and security \cite{sparrow1991application} \cite{carrington2011crime}. There are so many applications of centrality measure in different field network \cite{brohee2006evaluation} like psychological networks \cite{bringmann2019centrality} \cite{khojasteh2022comparing}, brain networks \cite{joyce2010new}, differential privacy models \cite{laeuchli2022analysis}, etc.
\section{Methods}\label{sec3}
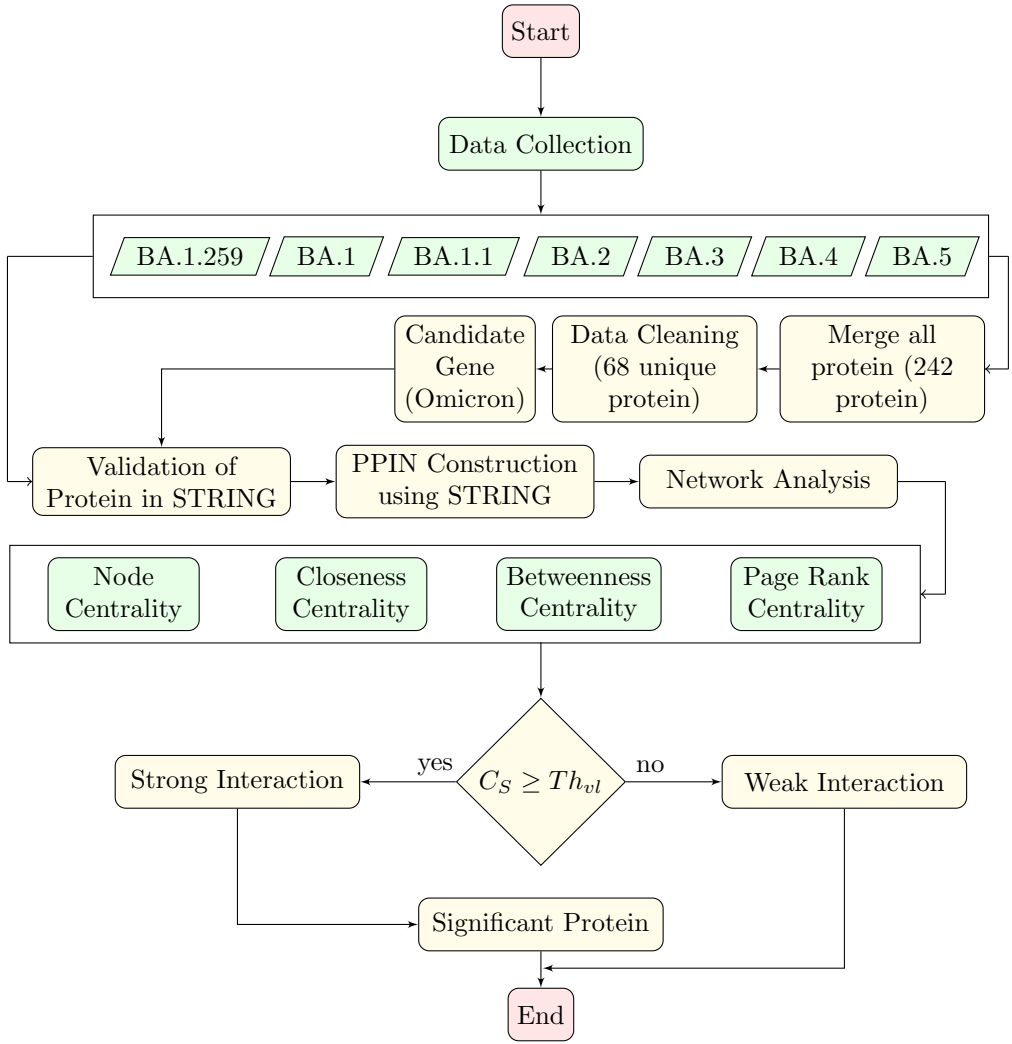
\begin{figure}
	\tikzstyle{startstop} = [rectangle, rounded corners, minimum width=.6cm, minimum height=.7cm,text centered, draw=black, fill=red!10]

	\tikzstyle{decision} = [diamond, draw, fill=yellow!10, text width=5.5em, text badly centered, inner sep=0pt]
	
	\tikzstyle{block} = [rectangle, draw, fill=yellow!10, 
	text width=9em, text centered, rounded corners, minimum height=2em,]
	
	\tikzstyle{block22} = [rectangle, draw, fill=yellow!10, 
	text width=8.5em, text centered, rounded corners, minimum height=2em,]
	
	\tikzstyle{line} = [draw, -latex']
	
	\tikzstyle{cloud} = [draw, ellipse,fill=red!20,
	minimum height=4em]
	
	\tikzstyle{decisionn} = [diamond, draw, fill=red!20, 
	text width=4em, text badly centered, node distance=4cm, inner sep=0pt]
	\tikzstyle{blockk} = [rectangle, draw, fill=blue!20, 
	text width=8em, text centered, rounded corners, minimum height=2em]
	\tikzstyle{blockkk} = [rectangle, draw, fill=red!20, 
	text width=6em, text centered, rounded corners, minimum height=2em]
	
	\tikzstyle{block1} = [rectangle, draw, 
	text width=7em, text centered, rounded corners, minimum height=2em, fill=green!10]

	\tikzstyle{blockkkk} = [rectangle, draw, 
	text width=5em, text centered, rounded corners, minimum height=2em, fill=green!10]
	
	\tikzstyle{blockkkkk} = [rectangle, draw, 
	text width=5.5em, text centered, rounded corners, minimum height=2em, fill=green!10]
	
	\tikzstyle{block2} = [rectangle, draw, 
	text width=4.6em, text centered, rounded corners, minimum height=2em, fill=yellow!10]
	
	\tikzstyle{block3} = [rectangle, draw, 
	text width=7em, text centered, rounded corners, minimum height=2em, fill=yellow!10]
	
	\tikzstyle{block4} = [rectangle, draw, 
	text width=7em, text centered, rounded corners, minimum height=2em, fill=yellow!10]
	
	\tikzstyle{process} = [rectangle, minimum width=11.8cm, minimum height=1.1cm, text centered, text width=3cm, draw=black, ]
	\tikzstyle{process2} = [rectangle, minimum width=12cm, minimum height=1.3cm, text centered, text width=3cm, draw=black, ]
	
	\tikzstyle{io} = [trapezium, trapezium left angle=70, trapezium right angle=110, minimum width=1.5cm, minimum height=.5cm, text centered, draw=black, fill=green!10]

	\begin{tikzpicture}[node distance = 1.5cm, auto]\label{ams1}
		
		\node [startstop] (start) {Start};
		\node [block1, below of=start, ] (datacollection) {Data Collection};
		\node [process, , below of=datacollection] (borobox) {};
		
		\node [left=10mm of borobox.west] (d1) {};
		\node [right=1.3mm of borobox.east] (d5) {};

		\node [io, , below of=datacollection, xshift=1.5em] (ba2) {BA.2 };
		\node [io,  , right of=ba2, ] (ba3) {BA.3};
		\node [io, , right of=ba3, ] (ba4) {BA.4};
		\node [io, , right of=ba4, ] (ba4) {BA.5};
		\node [io, , left of=ba2, xshift=-.5em] (ba11) {BA.1.1};
		\node [io, , left of=ba11, xshift=-.5em] (ba1) {BA.1};
		\node [io, , left of=ba1, xshift=-.3cm] (ba11259) {BA.1.259};

		\node [block2, below of=datacollection, xshift=-1cm, yshift=-1.5cm] (omicron) {Candidate Gene (Omicron)};
		\node [block3, right of=omicron, xshift=1cm] (datacleaning) {Data Cleaning (68 unique protein)};
		\node [block4, right of=datacleaning, xshift=1.5cm] (mergedata) {Merge all protein (242 protein)};
		\node [right=1.5mm of mergedata.east] (d6) {};
		\node [] at(d5 |- mergedata.east) (d6) {};
		\node [block, below of=omicron] (ppin) {PPIN Construction using STRING};
		\node [block, left of=ppin, xshift=-2.5cm] (validate) {Validation of Protein in STRING};
		\node [block, right of=ppin, xshift=2.5cm] (nwanalysis) {Network Analysis};

		\node [blockkkk, below of=validate,  xshift=-.5cm ] (nc) {Node Centrality};
		\node [] at(d1 |- validate.west) (d2) {};
		
		\node [blockkkk, right of=nc, xshift=1.5cm ] (cc) {Closeness Centrality};
		\node [blockkkkk, right of=cc, xshift=1.5cm ] (bc) {Betweenness Centrality};
		\node [blockkkk, right of=bc,  xshift=1.5cm] (pc) {Page Rank Centrality};

		\node [process2, below of=ppin] (boro2) {};
		\node [right=2mm of boro2.east] (d3) {};
		\node [right=3mm of nwanalysis.east] (d4) {};
			\node [] at(d3 |- nwanalysis.east) (d4) {};
	
		\node [decision, below of=datacollection, yshift=-7cm] (centrality) {$C_S \geq Th_{vl}$};
		
		\node [block22, right of=centrality, xshift=2.5cm] (weak) {Weak Interaction};
		\node [block22, left of=centrality, xshift=-2.5cm] (strong) {Strong Interaction};
		\node [block22, below of=centrality, yshift=-.4cm] (improtein) {Significant  Protein};
		\node [startstop, below of = improtein, yshift=.3cm] (end) {End};
		
		\node [above=3.4cm of improtein] (d8) {};
		\node [below=1mm of improtein, xshift=-1mm] (d10) {};

		\path [line] (weak) |- (d10);
		\path [line] (start) -- (datacollection);
		\path [line] (datacollection) -- (borobox);
		\path [line] (mergedata) -- (datacleaning);
		\path [line] (datacleaning) -- (omicron);
		\path [line] (validate) -- (ppin);
		\path [line] (ppin) -- (nwanalysis);
		\path [line] (omicron) -| (validate);
		\path [line] (d8) --(centrality.north);
		\path [line] (centrality) -- node[anchor=west, yshift=2mm] {yes}(strong);
		\path [line] (centrality) -- node[anchor=east, yshift=2mm] {no}(weak);
		\path [line] (strong) |- (improtein);
		\path [line] (improtein) -- (end);
		
		\draw [] (borobox) -- (d1.center);
		\draw [] (d1.center) -- (d2.center);
		\draw [->] (d2.center) -- (validate.west);
		
		\draw [] (nwanalysis) -- (d4.center);
		\draw [] (d4.center) -- (d3.center);
		\draw [->] (d3.center) -- (boro2.east);
		
		\draw [] (borobox) -- (d5.center);
		\draw [] (d5.center) -- (d6.center);
		\draw [->] (d6.center) -- (mergedata.east);
		
		
	\end{tikzpicture}
	\caption{Flow chart for numerical solution procedure}\label{flow}
\end{figure}

This study used Omicron lineage variants data. The research has been completed in different steps like, data collection, data cleaning, data validation, creation of PPIN data, centrality measure and finally clustered the whole network in different clusters. The clustering is done by the MCL (Markov clustering algorithm) \cite{satuluri2009scalable}. The objective of this research work is to get the significant protein or prioritize the protein. For this we have focuses on the centrality measure of the network. Figure \ref{flow} illustrates the research workflow. 
\subsection{Data Collection} \label{subsec1}
We have taken the real dataset of Omicron from Universal Protein Resource/SwissProt (UniProt/ SwissProt) \cite{uniprot2021uniprot} database which is reviewed and found in the human body. In addition to storing experimental results, computational features, and scientific conclusions, Swiss-Prot is a highly annotated, non-redundant protein sequence database. Currently, the UniProt Knowledgebase is comprised of UniProtKB/Swiss-Prot, which has been reviewed. It provides accurate, consistent, and rich annotations for functional information about proteins. Initially, we have taken a total of 228 proteins: B.1.1.529 (27), BA.5 (30), BA.4 (31), BA.3 (34), BA.2 (38), BA.1.1 (34), BA.1 (34) and analyzed individual Omicron lineage PPIN. The PPIN of Omicron Lineage Variants are shown in Figure \ref{fig:B11529} to Figure \ref{fig:BA5}. Then we sum up the data and cleaned the data by removing duplicate data entries to create the Omicron PPIN. The data validation and PPIN data creation in all the cases are done by STRING \cite{szklarczyk2015string}. There are several sources of information within the STRING database, including computational prediction methods, experimental data, and public text collections. A regular update keeps it up-to-date and it is free to access. Additionally, it generates network images using a spring model. In this model, nodes are considered masses, and edges are considered springs. After cleaning the data we gate unique 68 proteins which create the Omicron PPIN. 
\subsection{Centrality Measure} \label{subsec2}
\begin{figure}[ht]
	\begin{center}
	\begin{minipage}[b]{0.45\linewidth}
		\hspace{-1.4cm}
		\includegraphics[ width=9cm, height=5cm]{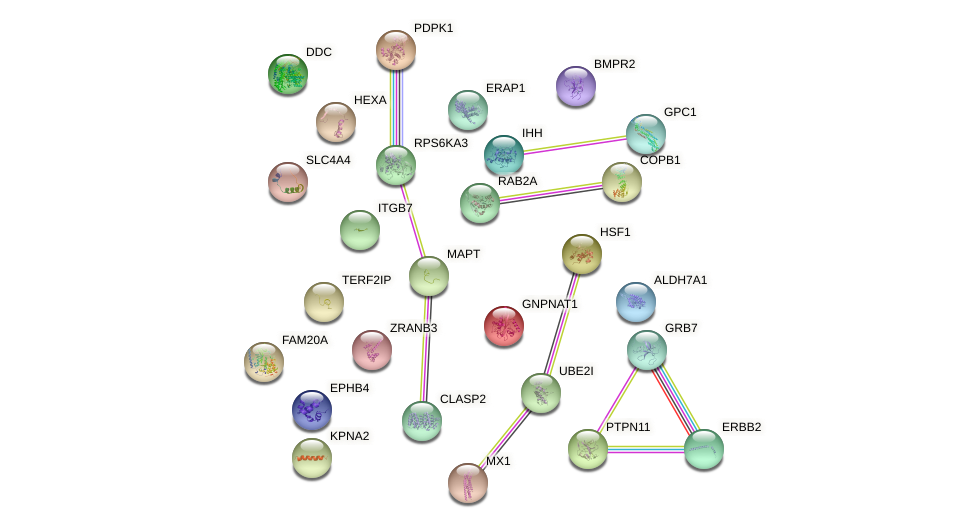}
		\caption{PPIN of \mbox{B.1.1.529}}
		\label{fig:B11529}
	\end{minipage}\hfill
	\begin{minipage}[b]{0.45\linewidth}
		\centering
		\includegraphics[width=8cm, height=5cm]{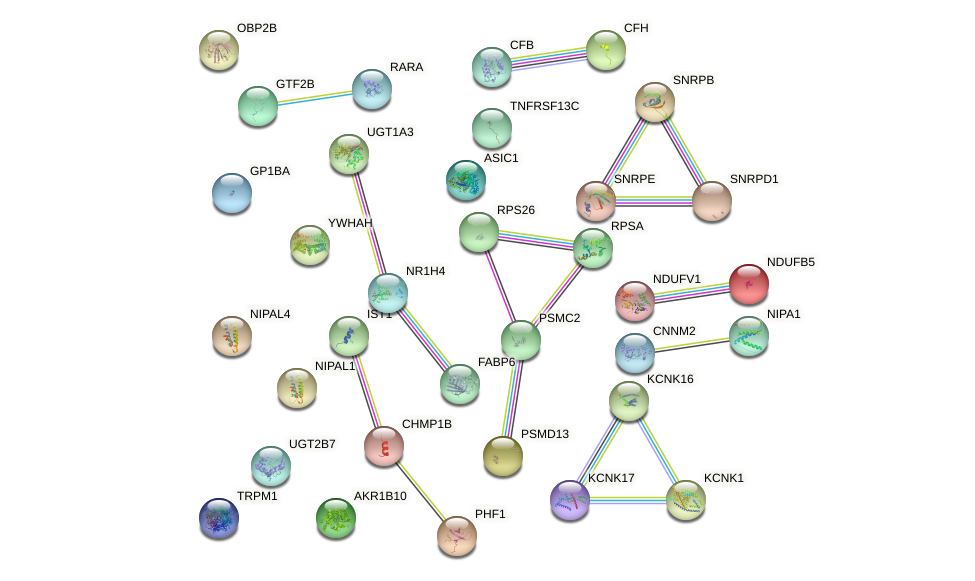}
		\caption{PPIN of \mbox{BA.1}}
		\label{fig:BA1}
	\end{minipage}\hfill \vspace{1cm}
	\begin{minipage}[b]{0.45\linewidth}
		\centering
		\includegraphics[width=6cm, height=5cm]{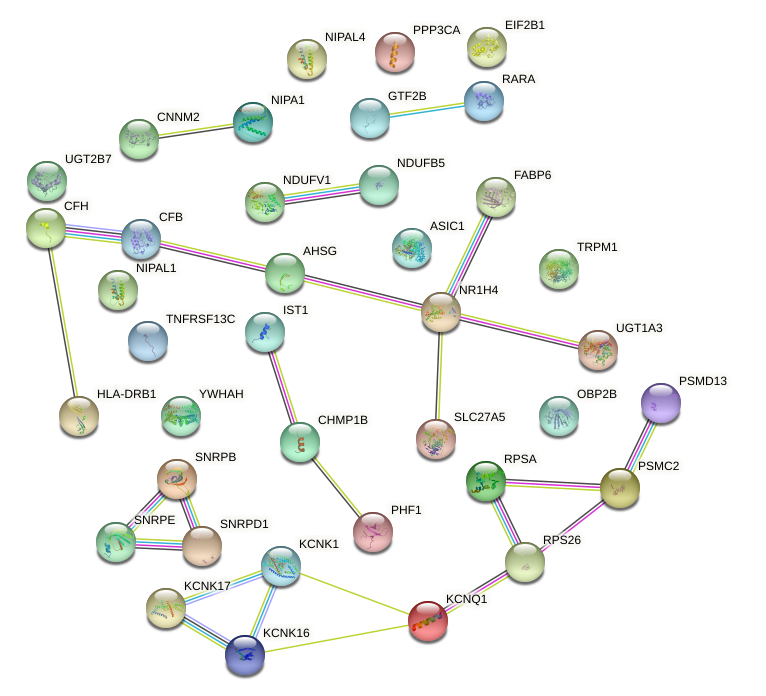}
		\caption{PPIN of \mbox{BA.2}}
		\label{fig:BA2}
	\end{minipage}\hfill
	\begin{minipage}[b]{0.45\linewidth}
		\centering
		\includegraphics[width=8cm, height=5cm]{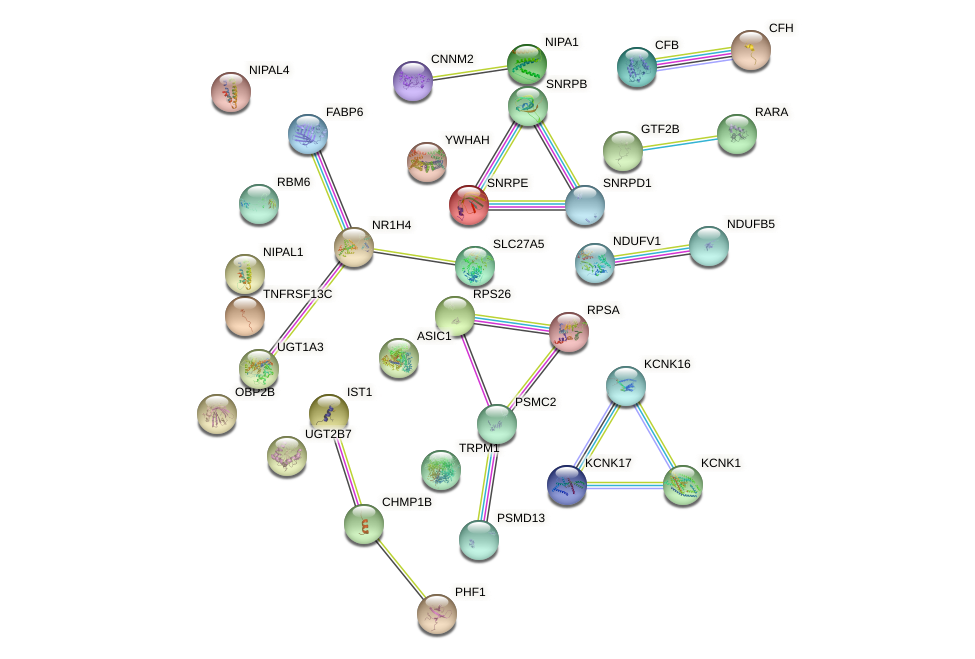}
		\caption{PPIN of \mbox{BA.3}}
		\label{fig:BA3}
	\end{minipage}\hfill \vspace{1cm}
	\begin{minipage}[b]{0.45\linewidth}
		\hspace{-.8cm}
		\includegraphics[width=8cm, height=5cm]{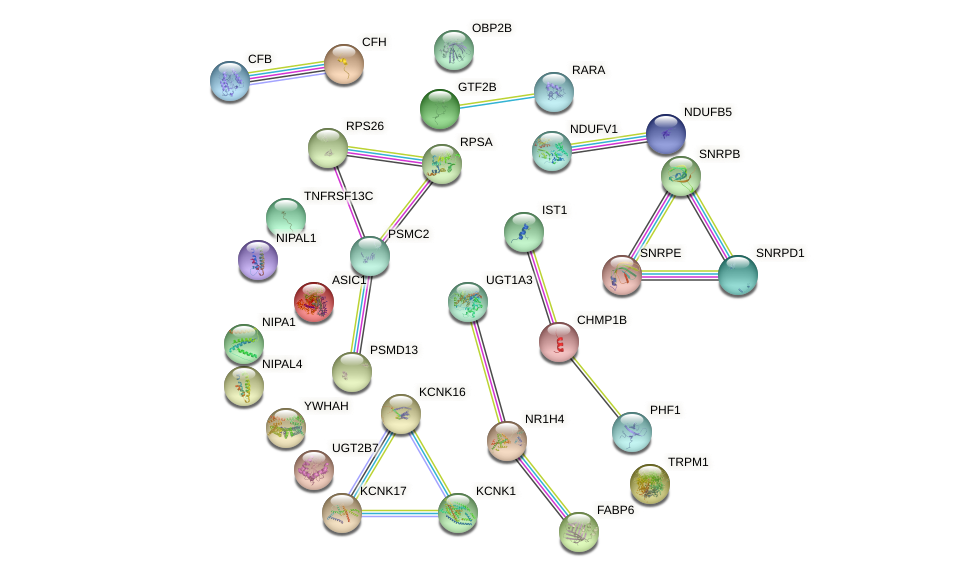}
		\caption{PPIN of \mbox{BA.4}}
		\label{fig:BA4}
	\end{minipage}\hfill
	\begin{minipage}[b]{0.45\linewidth}
		\centering
		\includegraphics[width=8cm, height=5cm]{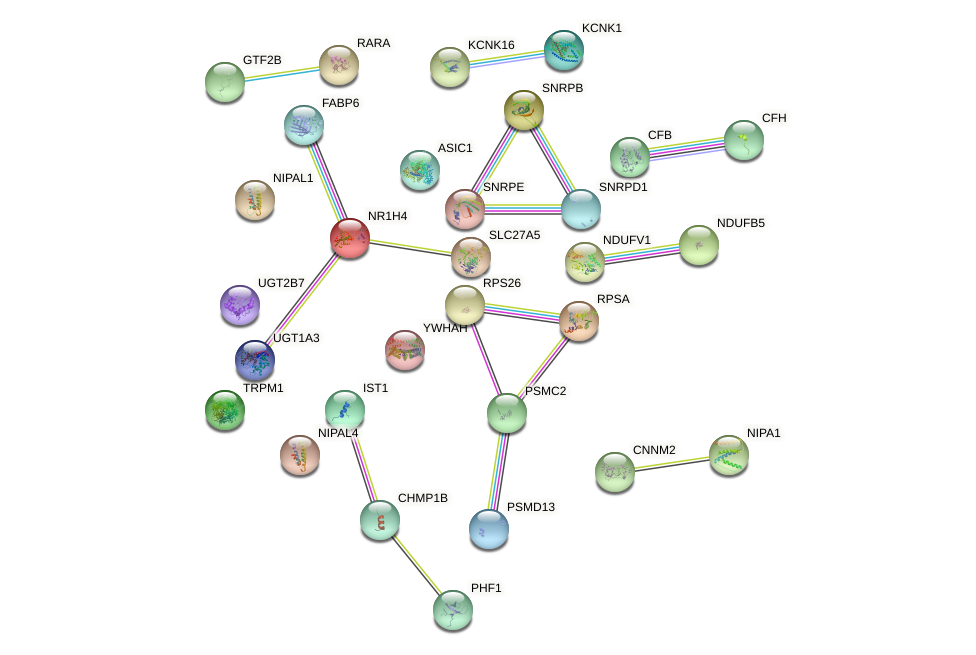}
		\caption{PPIN of \mbox{BA.5}}
		\label{fig:BA5}
	\end{minipage}
	\end{center}
\end{figure}

Here we will discuss very interasting aspect of network measure called centrality. Centrality is basically widely used measure of how central a particular node is with respect to the network.The network that results from the PPI data is thought to be an undirected graph. Each node's weight in the graph is determined by the centrality approach. The BC, CC, DC, and PRC are a few centrality techniques that can be applied to undirected graphs. Figure \ref{fig:B11529} to \ref{fig:BA5} and \ref{fig:omnicron} depicts a protein network as an example of an undirected graph. The variant BA.1 and BA.1.1 has the same PPIN only the difference in mutation. The edges of the graph reflect the functional interaction or relationship that takes place between proteins, whereas the nodes in the graph demonstrate the proteins that affect Omicron's activity.
\subsubsection{Degree Centrality}
The first basic centrality measure is the degree centrality (DC) \cite{freeman1977set}. We know that the degree is basically the number of edges which are adjacent on a particular node. The DC is esentially is a degree of a node but it is normalized. 

The DC of a node $\mathit{v}$ is a degree of the node $\mathit{v}$ and divided by the maximum degree of a node present in the graph. 
A node's degree centrality $C_d\mathit{(v)}$ in a network G(V, E) is denoted mathematically as follows:

\begin{equation}
	C_d\mathit(v) = \frac{deg\mathit(v)}{\mbox{max deg}_{u \in \mathit{v}} \mathit(u)}  .\label{eq1}
\end{equation}

It basically ranges between $0$ to $1$ and more the degree centrality mean higher the likelyhood that the node has maximum degree. The $C_d\mathit(v)$ can use to identify the more prominent or influential node from a network. 
\subsubsection{Closensee Centrality}
The Closensee Centrality ($CC$) \cite{sabidussi1966centrality} indicates how close a node from the rest of the network. A approach to identify nodes that can efficiently spread information throughout a graph is through their CC. Average distance between a node and all other nodes is measured by its proximity centrality. The distances between nodes that have a high proximity score are the shortest. A node's Closeness centrality $C_c(\mathit{v})$ in a graph G(V, E) is denoted mathematically as follows:  
\begin{equation}
	C_c\mathit(v) = \frac{\lvert V \rvert  - 1}{\sum_{u\in V - {\mathit{\{ v\}}}}^{}d\mathit (u, v)}  .\label{eq2}
\end{equation}\vspace{3mm}

where, number of nodes is given by $ \lvert V \rvert $ and the distance between two nodes $\mathit{u}$ and $\mathit{v}$ is represented as $\mathit{d(u, v)}$. Higher the value of $CC$ better would be the quality of the particular node. The measure is useful in examining or restricting the spreds of disease in epidemic modelling.

\subsubsection{Betweenness Centrality}
Betweenness (BC) \cite{freeman1977set} is the measure to compute how central a node is in between paths of the network or we can say to compute how many paths(shortest) of the network passes through the node. A node's Betweenness centrality $C_b(\mathit{v})$ in a network G(V, E) is denoted mathematically as follows: 
\begin{equation}
	C_b\mathit(v) =\sum_{xy \in V - \mathit{\{v\}}}^{} \frac{\sigma_{xy} (\mathit{v})}{\sigma_{xy}}
\end{equation}

where the frequency of shortest paths in the network between nodes $x$ and $y$ is indicated by $\sigma_{xy}$ and $\sigma_{xy} (\mathit{v})$ denotes the same passing through $\mathit{v}$. If $x= 1$, then $\sigma_{xy} = 1$. The BC is useful in identifying the super spreaders in analyzing disease spreading in epidemiology.

\subsubsection{Page Rank}
PageRank centrality \cite{ivan2011web} is an adaptation of Eigen centrality that ranks web content by using the value of linkages between sites. Any type of network, including protein interaction networks, can be used with it. Mathematically, the pagerank Centrality $C_PR(\mathit{v})$ in a network $G(V, E)$ of a node $\mathit{v_i}$ is defined as:
\begin{equation}
		C_{PR}\mathit(v_i) = \frac{1-d}{\vert V \vert} + d \sum_{\mathit(v_t)\in Inneighbor\mathit(v_i)}^{} \frac{C_{PR}(\mathit{v_t})}{outdeg(\mathit{v_t})}
\end{equation} \vspace{3mm}

Where $d$ is constant and called damoing factor, usually the constant value is considered as $0.85$.
\subsection{Markov Clustering}
At the Centre for Mathematics and Computer Science in the Netherlands, Stijn van Dongen created the Markov Cluster Algorithm, MCL algorithm \cite{vandongen2000cluster}. It is an unsupervised cluster approach for networks that is extremely quick and scalable and is based on the simulation of graph flow. It is employed in bioinformatics and other fields. The distance matrix derived from the STRING global scores in our study serves as the input to MCL. Higher global scores for these interacting proteins increase the likelihood that they will cluster together. The MCL \cite{das2021markov}\cite{das2022analytical} operates primarily in two ways: Expansion-the operation corresponds to the multiplication of standard matrices and simulates how a flow spreads and becomes more homogeneous. The next is inflation which is described logically as a diagonal scaling proceeded by a Hadamard power. Flow is compressed by inflation by thickening only in areas where current density is high and thinning only in areas where current density is low.  There is no way to know how many clusters there are. With the help of the inflation parameter, it is implicitly managed. Higher inflation results in more clusters being obtained, which is indirectly connected to the clustering's precision. Here, the inflation value has been set at $2$.

\section{Results and Discussion}\label{sec4}
\begin{figure}[ht]
	\begin{center}
			\begin{minipage}[b]{0.45\linewidth}
			\centering
			\includegraphics[width=\textwidth]{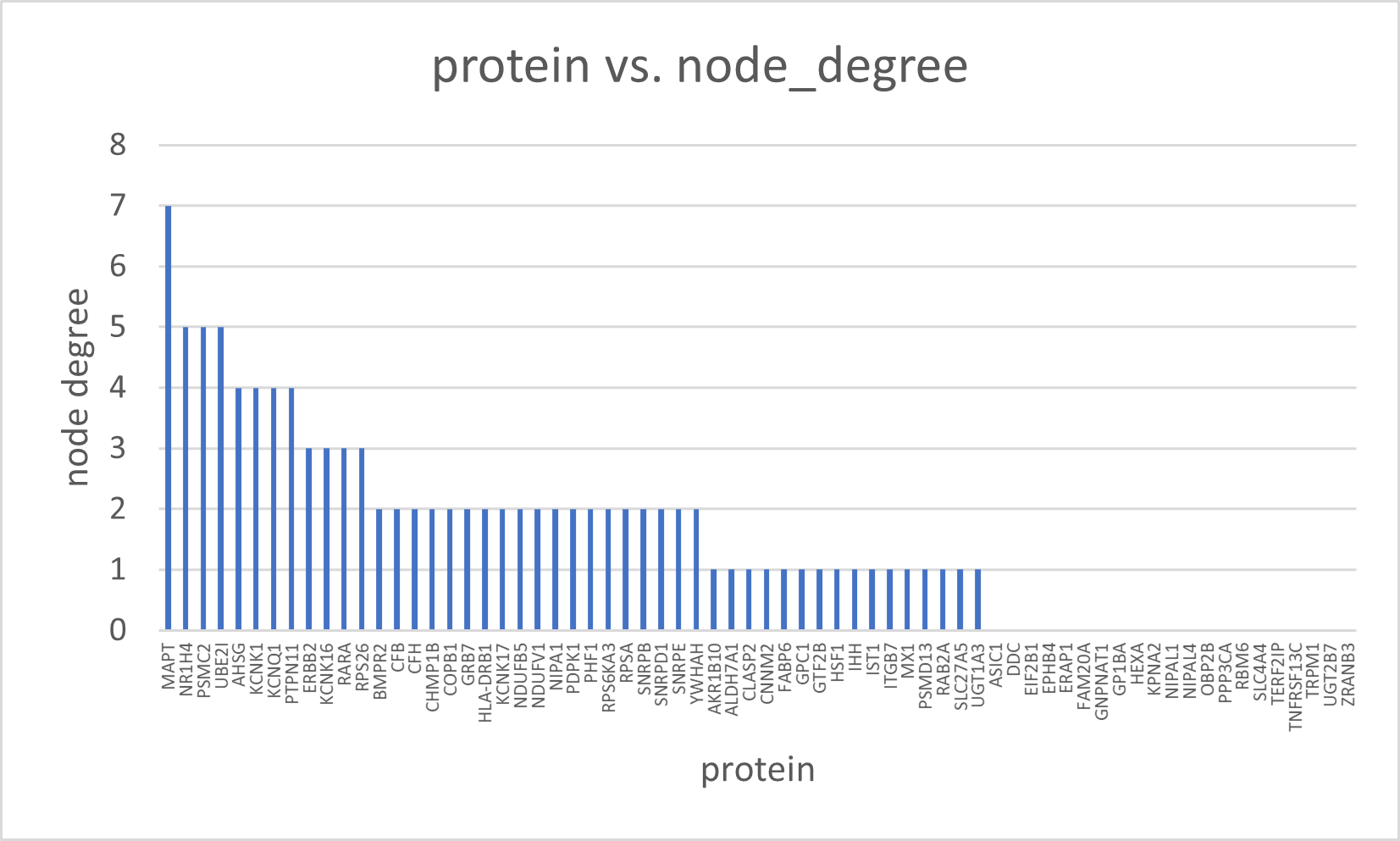}
			\caption{protein vs. node degree}
			\label{fig:degree}
		\end{minipage}
		\begin{minipage}[b]{0.45\linewidth}
			\centering
			\includegraphics[width=\textwidth]{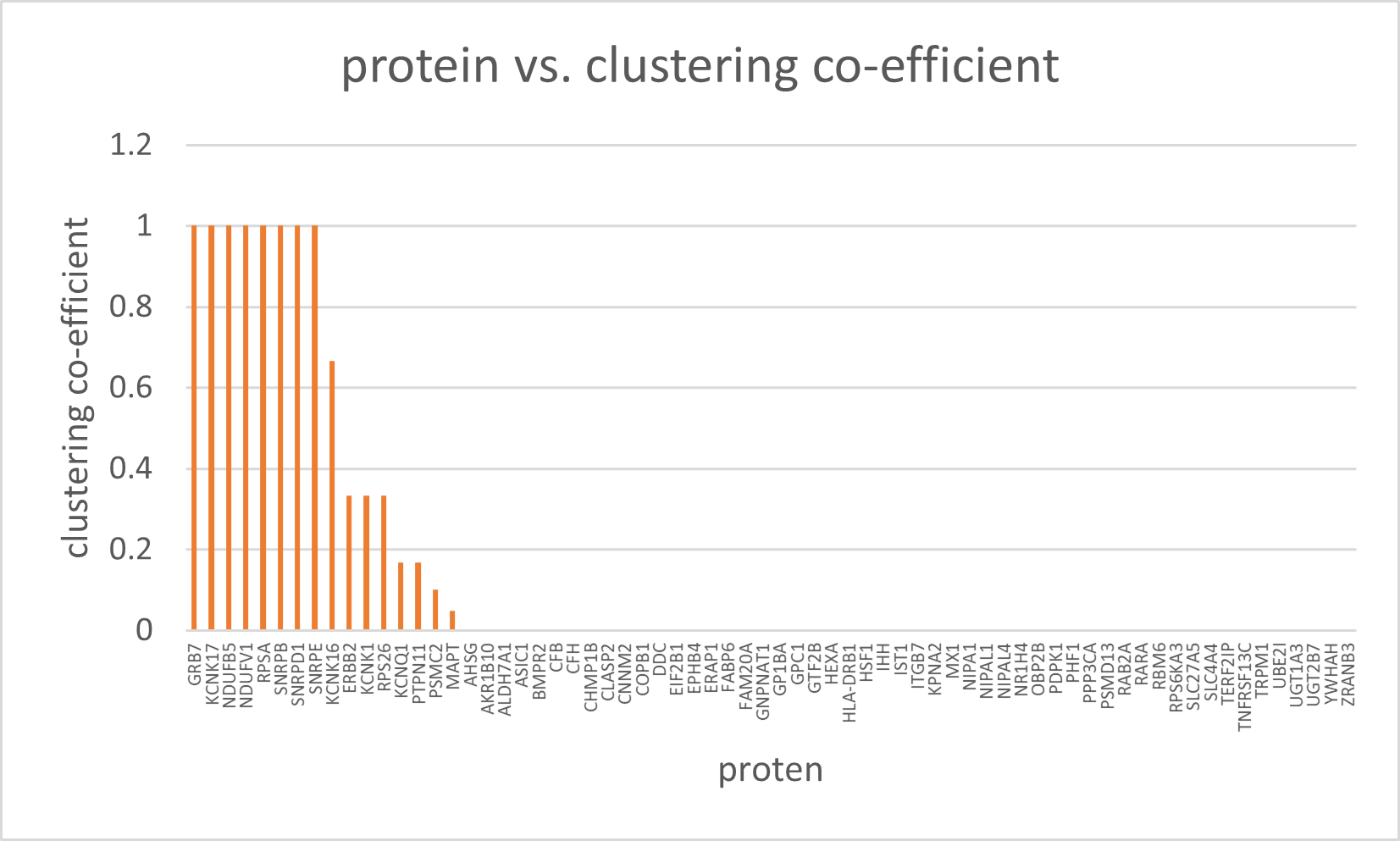}
			\caption{protein vs. CCo}
			\label{fig:CCo}
		\end{minipage}
		\begin{minipage}[b]{0.45\linewidth}
			\centering
			\includegraphics[width=\textwidth]{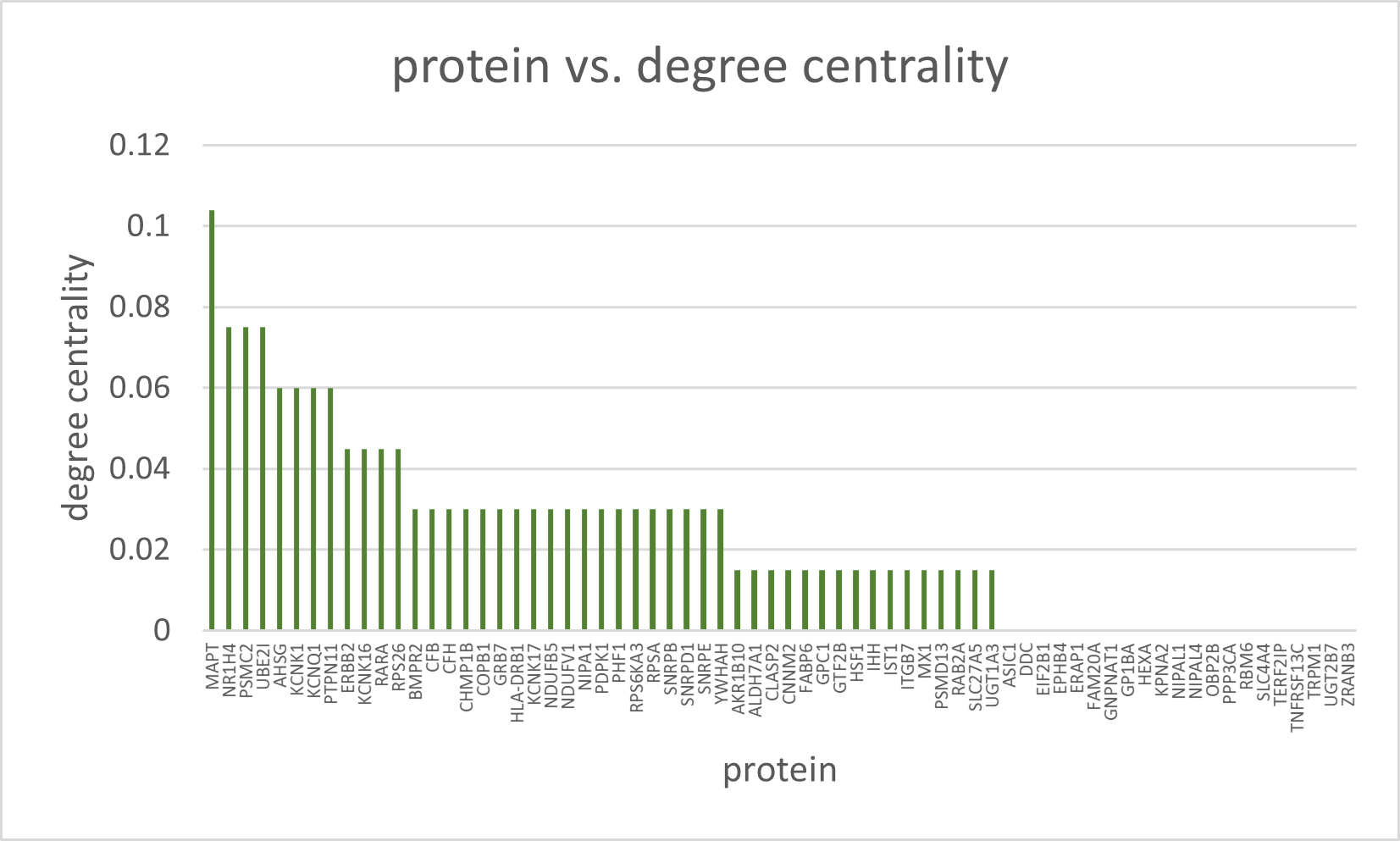}
			\caption{protein vs. DC}
			\label{fig:degree_c}
		\end{minipage}
		\begin{minipage}[b]{0.45\linewidth}
			\centering
			\includegraphics[width=\textwidth]{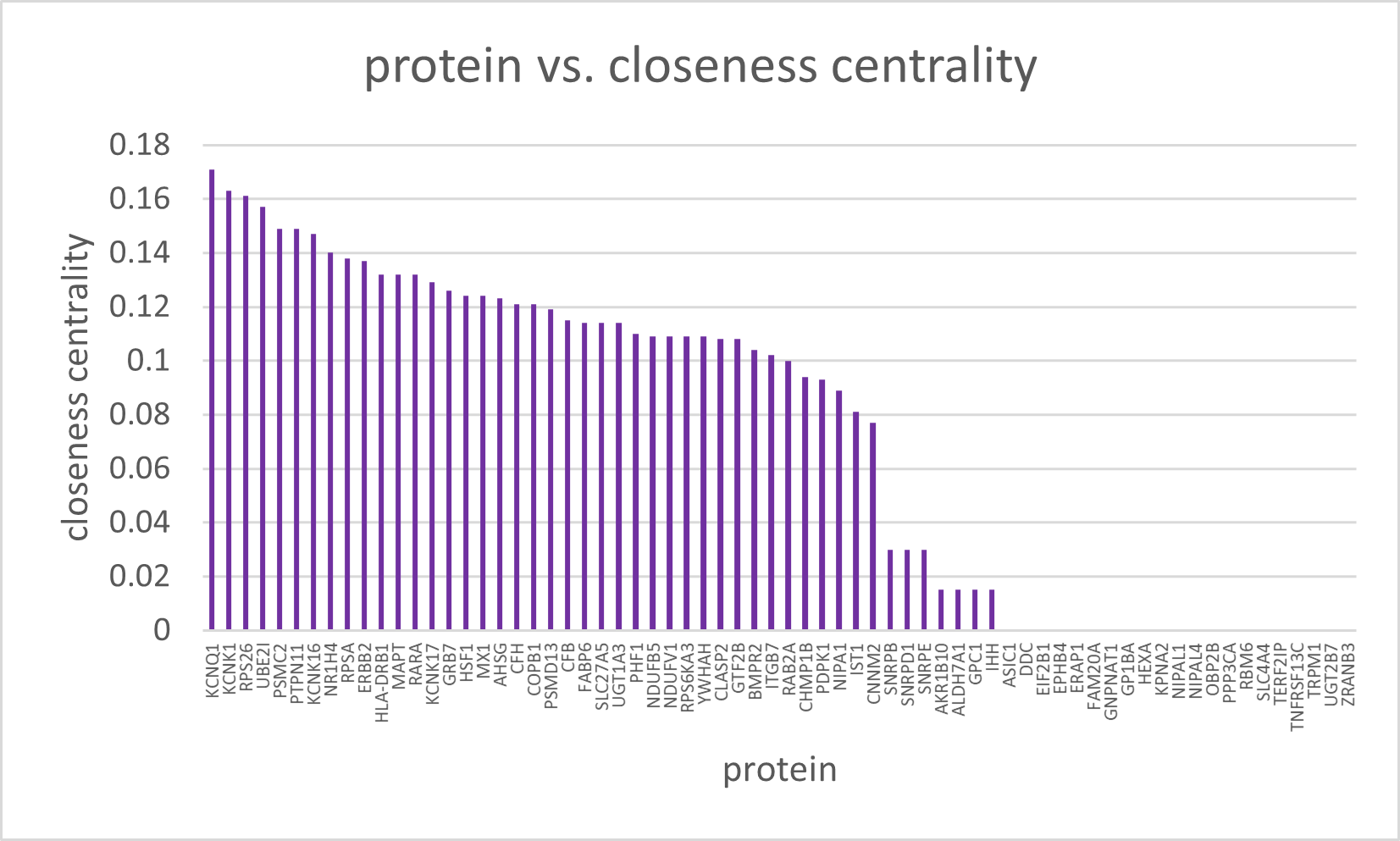}
			\caption{protein vs. CC}
			\label{fig:closeness_c}
		\end{minipage}
		\begin{minipage}[b]{0.45\linewidth}
			\centering
			\includegraphics[width=\textwidth]{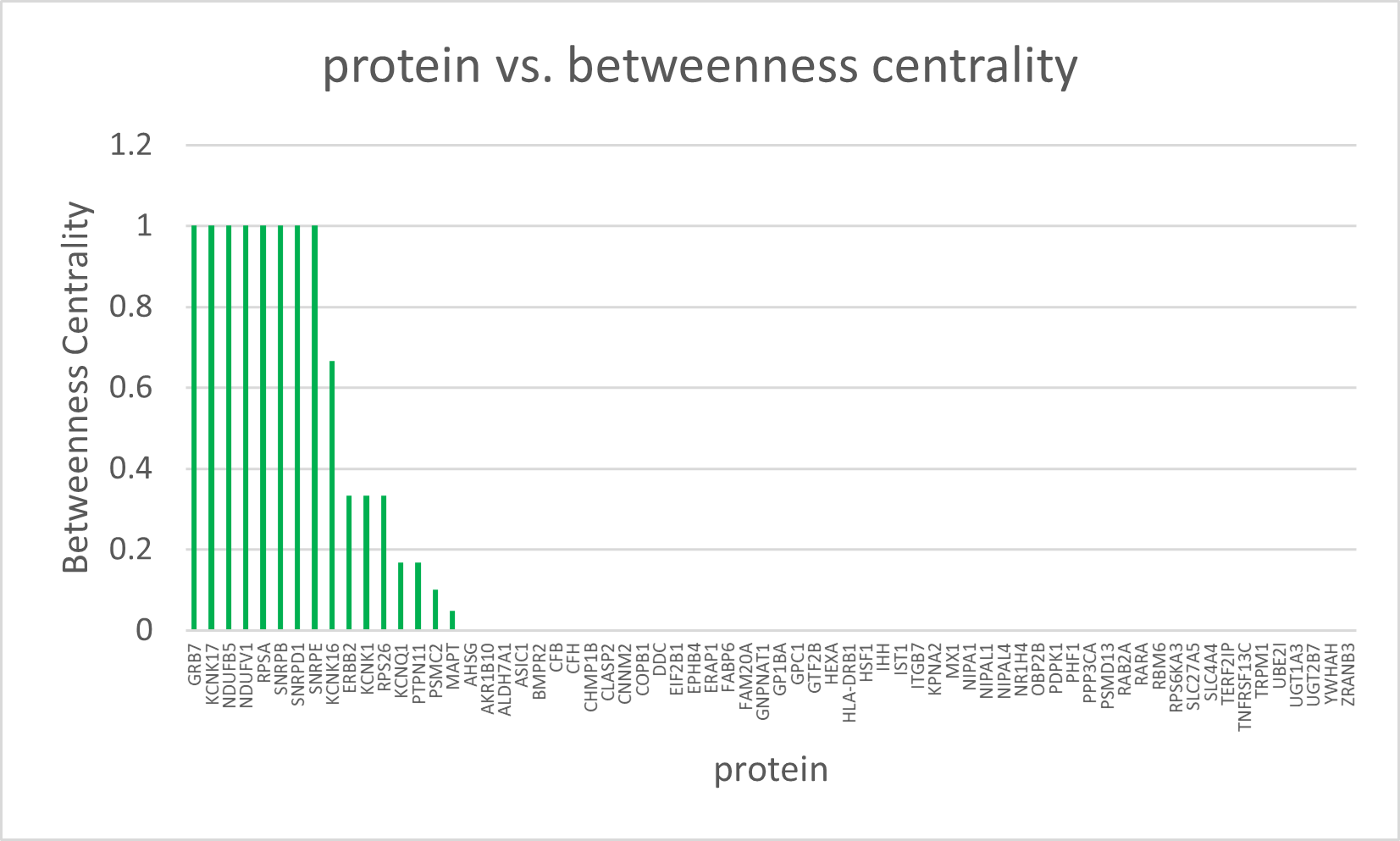}
			\caption{protein vs. BC}
			\label{fig:betweenness_c}
		\end{minipage}
		\begin{minipage}[b]{0.45\linewidth}
			\centering
			\includegraphics[width=\textwidth]{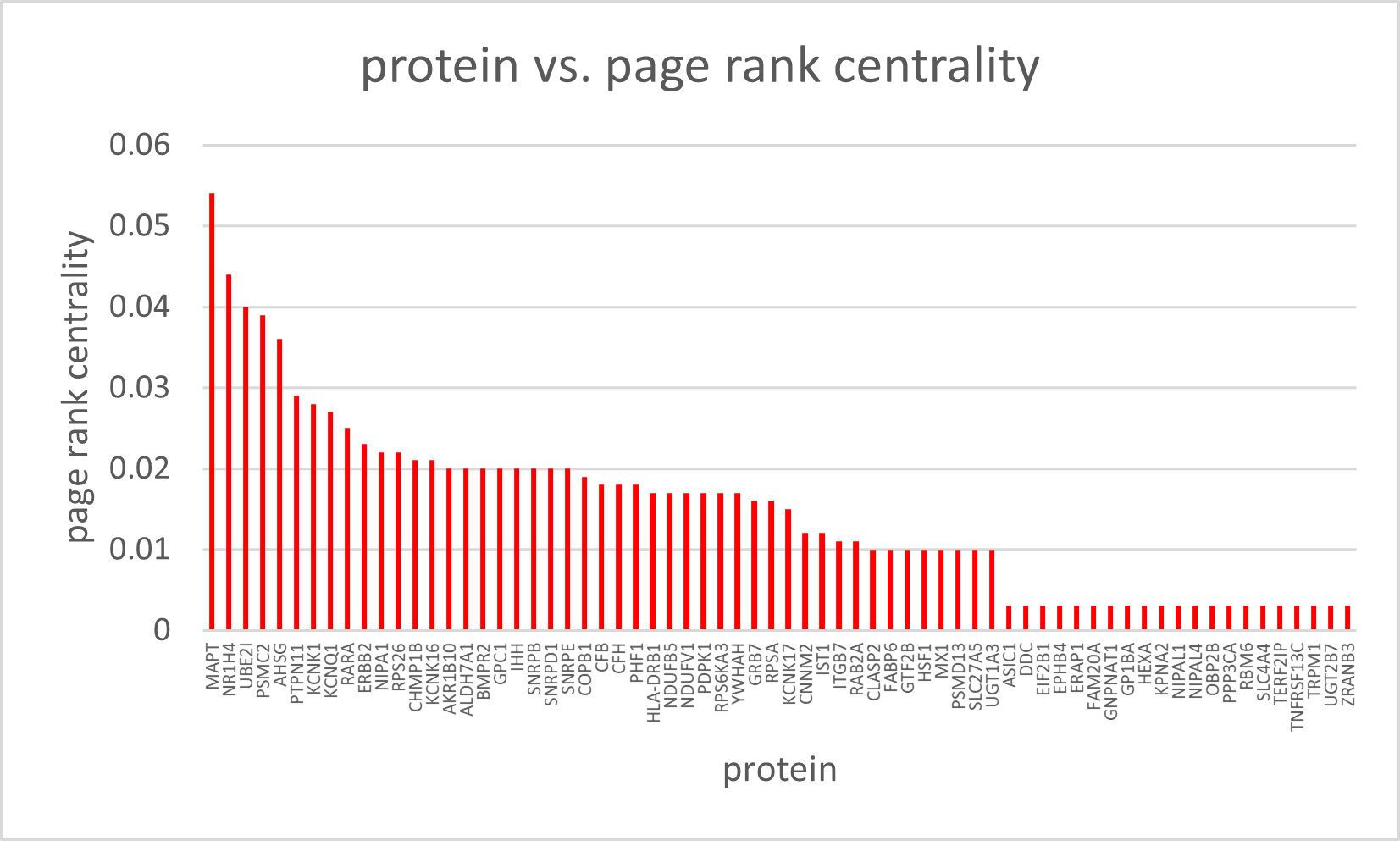}
			\caption{protein vs. PR}
			\label{fig:pagerank_c}
		\end{minipage}
		\begin{minipage}[b]{0.45\linewidth}
			\centering
			\includegraphics[width=6cm, height=9cm]{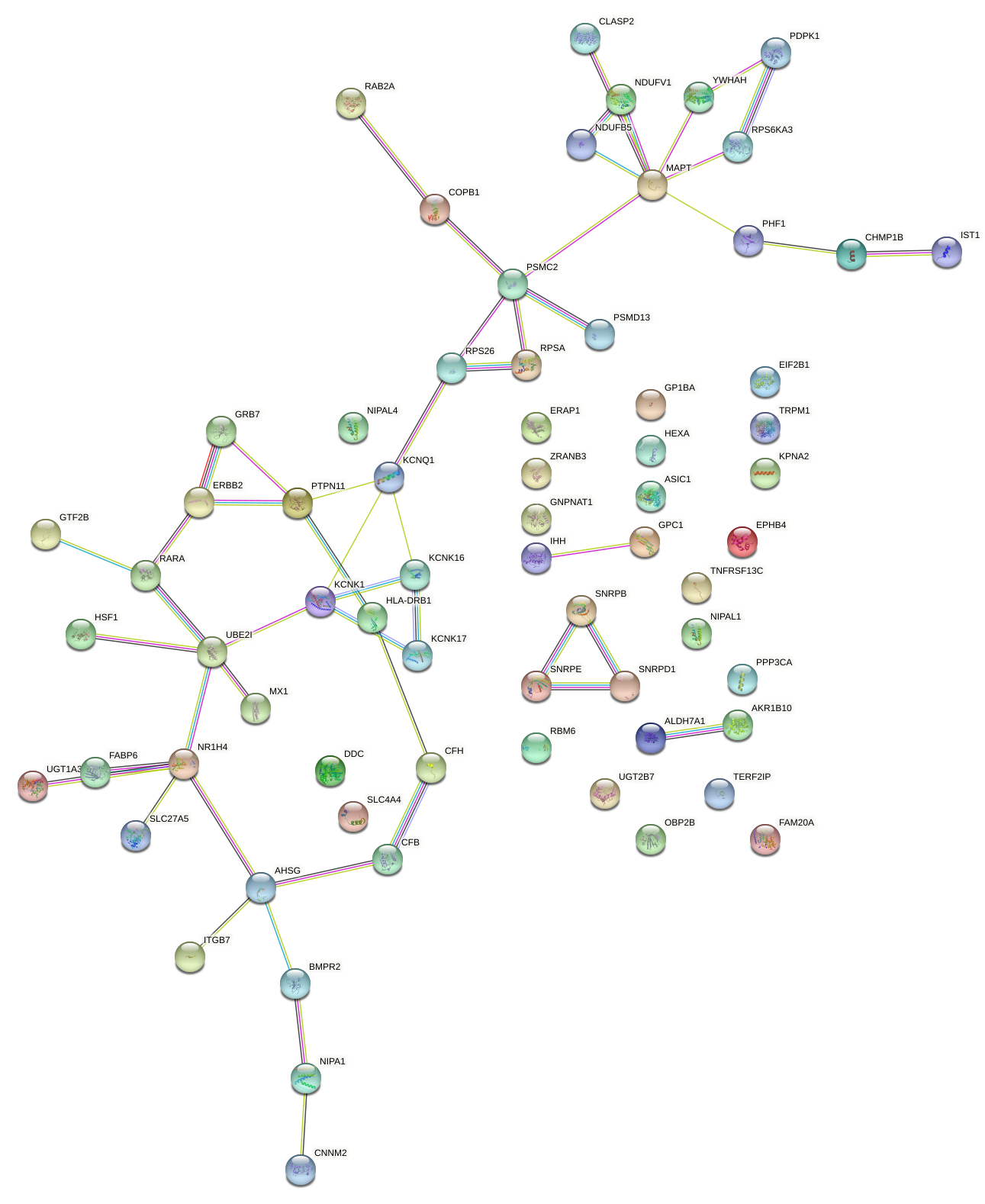}%
			\caption{Omnicron PPI network}
			\label{fig:omnicron}
		\end{minipage}
		\begin{minipage}[b]{0.45\linewidth}
			\centering
			\includegraphics[width=6cm, height=9cm]{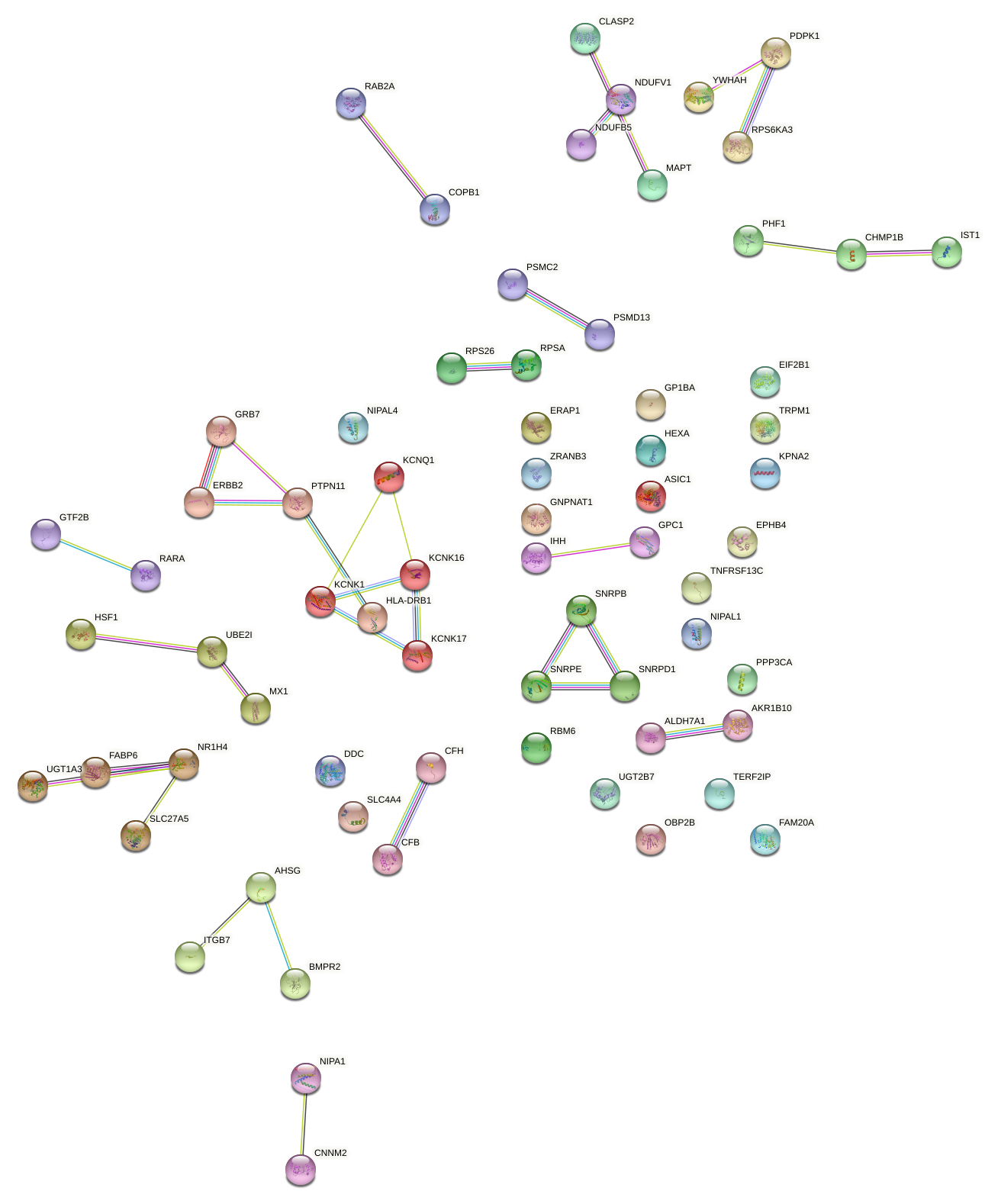}%
			\caption{PPIN after clustering}
			\label{fig:mcl}
		\end{minipage}\hfill
	\begin{minipage}{350pt}

\end{minipage}
\end{center}
\end{figure}

The global properties of Omicron base lineage variants are shown in \ref{tab4}. All the seven network except $BA.1.1.259$ has an average node degree greater than $1$. The $3$ base lineage ($BA.1.1.259, BA.1, BA.1.1$) has same density $0.0284$. The highest density is $0.06719$ (BA.4) and the lowest density belongs to BA.3 ($0.00416$). The average LCC is pretty good (highest $0.771$). The best network is the BA.2 with the smallest p-value($0.00038$).
The Table \ref{tab1} shows the global  features of the Omicron PPIN. Node degree is $1.53$ on average and the density is $0.0228$. The information in Table \ref{tab2} contains the centrality scores of $68$ proteins, which allow us to identify the protein's relevance. The network has a maximum degree of $7$ with an average Local Clustering Coefficient (LCC) of $0.385$. The LCC range from $0$ to $1$, and they represent the density of connections among neighbours. Nodes that have higher values belong to densely connected clusters. The node is considered a part of the clique if it has a value of $1$. The proteins GRB7, KCNK17, NDUFB5, NDUFV1, RPSA, SNRPB, SNRPD1, and SNRPE in Table \ref{tab2} are containing CCo value as $1$ as they are part of the clique. Figure \ref{fig:omnicron} is showing the PPI network of Omicron and the score of the CM are visualizing in Figure \ref{fig:degree} to \ref{fig:pagerank_c}. We have calculated the maximum value of each centrality measure and divided it by two to get each category's threshold value. The threshold value will help us to signify the important protein in the network. We have highlighted the significant protein by getting the intersection of all the important proteins of each category (CC, DC, and, PCR). A total of $8$ significant proteins were detected from $68$ unique proteins. In our research work, we have extracted the $18$ network cluster from the Omicron main network with the help of the Markov clustering algorithm shown in Figure \ref{fig:omnicron} and \ref{fig:mcl}. In Table \ref{tab7} we can see, cluster $\mathcal{C}_1, \mathcal{C}_2$ and $\mathcal{C}_3$ has $4$ protein in each, $\mathcal{C}_4$ to $\mathcal{C}_8$ has $4$ protein in each and rest of the clusters are containing $2$ protein in each.

\begin{table}[ht]
	\begin{center}
		\begin{minipage}{350pt}
			\caption{Global properties of Omicron lineage variant's network}\label{tab4}%
			\begin{tabular}{@{}llllllll@{}}
				\toprule
				Variants & \#node & \#edge & avg. node degree & max degree & density  & avg. LCC & P-value  \\
				\midrule
				B.1.1.529 & 27 & 10 & 0.741 & 2 & 0.0284 & 0.407 & 0.3370 \\  \vspace{.5mm}
				BA.1 & 34 & 18 & 1.06 & 3 & 0.0284 & 0.627 & 0.00365\\\vspace{.5mm}
				BA.1.1& 34 & 18 & 1.06 & 3 & 0.0284 & 0.627 & 0.00365\\\vspace{.5mm}
				BA.2 & 38 & 25 & 1.32 & 4 & 0.03556  & 0.535 & 0.00038\\\vspace{.5mm}
				BA.3 & 34 & 19 & 1.12 & 3 & 0.00416  & 0.657 & 0.00104\\\vspace{.5mm}
				BA.4 & 31 & 17 & 1.1 & 3 & 0.06719  & 0.624 & 0.00271\\\vspace{.5mm}
				BA.5 & 30 & 17 & 1.13 & 3 & 0.05666  & 0.711 & 0.00236\\
				\botrule
			\end{tabular}
			\footnotetext{avg. LCC: Average Local clustering coefficient, P-value: PPI enrichment P-value}
		\end{minipage}\hfill \vspace{.6cm}
		\begin{minipage}{350pt}
				\caption{Global properties of Omicron network}\label{tab1}%
				\begin{tabular}{@{}llllllll@{}}
					\toprule
					
					\#Node & \#Edges & Max Degree & Avg. node degree & Density & Avg. LCC & p-value\\
					\midrule
					68 & 52 & 7 & 1.53 & 0.0228 & 0.385 & 0.0963 \\ 
					\botrule
				\end{tabular}
				\footnotetext[1]{\tiny Avg LCC: Avg. Local Clustering Coefficient, p-value: PPI enrichment p-value}
			\end{minipage}
	\end{center}
\end{table}
\vspace{.5cm}
\begin{table}[ht]
	\begin{center}\caption{Generated $18$ clusters from MCL algorithm}\label{tab7}%
		\begin{tabular}{@{}llllll@{}}
			\toprule
			cluster & gene count & protein names  \\
			\midrule
$\mathcal{C}_1$ &4 & KCNK1,KCNK16,KCNK17,KCNQ1 \\ \vspace{2mm}
$\mathcal{C}_2$ &4 & ERBB2,GRB7,HLA-DRB1,PTPN11 \\ \vspace{2mm}
$\mathcal{C}_3$ &4 & FABP6,NR1H4,SLC27A5,UGT1A3\\ \vspace{2mm}
$\mathcal{C}_4$ &3 & PDPK1,RPS6KA3,YWHAH \\ \vspace{2mm}
$\mathcal{C}_5$ &3 & HSF1,MX1,UBE2I \\ \vspace{2mm}
$\mathcal{C}_6$ &3 & AHSG,BMPR2,ITGB7\\ \vspace{2mm}
$\mathcal{C}_7$ &  3 & SNRPB,SNRPD1,SNRPE\\  \vspace{2mm}
$\mathcal{C}_8$ &3 & CHMP1B,IST1,PHF1\\ \vspace{2mm}
$\mathcal{C}_9$ &2 & RPS26,RPSA \\ \vspace{2mm}
$\mathcal{C}_{10}$ &2 & CLASP2,MAPT\\ \vspace{2mm}
$\mathcal{C}_{11}$ &2 & COPB1,RAB2A\\ \vspace{2mm}
$\mathcal{C}_{12}$ &2 & PSMC2,PSMD13\\ \vspace{2mm}
$\mathcal{C}_{13}$ &2 & GTF2B,RARA\\  \vspace{2mm}
$\mathcal{C}_{14}$ &2 & NDUFB5,NDUFV1\\ \vspace{2mm}
$\mathcal{C}_{15}$ &2 & GPC1,IHH\\ \vspace{2mm}
$\mathcal{C}_{16}$ &2 & CNNM2,NIPA1\\ \vspace{2mm}
$\mathcal{C}_{17}$ &2 & AKR1B10,ALDH7A1\\ \vspace{2mm}
$\mathcal{C}_{18}$ &2 & CFB,CFH \\

			\botrule
		\end{tabular}
	\end{center}
\end{table}

\begin{table}[!h]
	\begin{center}
		
		\begin{minipage}{250pt}
			\caption{Centrality measure and some important score of 68 Omicron protein}\label{tab2}%
			\setlength\extrarowheight{-3pt}
			\begin{tabular}{@{}llllllll@{}}
				\toprule

				S.N.& Protein Name & Node Degree & CCo & DC & CC & BC & PRC \\
				\midrule
				
				\tiny	 1	 	&	\tiny	 	AHSG	 	&	\tiny	 	4	 	&	\tiny	 	0	 	&	\tiny	 	0.06	 	&	\tiny	 	0.123	 	&	\tiny	 	0	 	&	\tiny	 	0.036	\\ 
				\tiny	 2	 	&	\tiny	 	AKR1B10	 	&	\tiny	 	1	 	&	\tiny	 	0	 	&	\tiny	 	0.015	 	&	\tiny	 	0.015	 	&	\tiny	 	0	 	&	\tiny	 	0.02	\\ 
				\tiny	 3	 	&	\tiny	 	ALDH7A1	 	&	\tiny	 	1	 	&	\tiny	 	0	 	&	\tiny	 	0.015	 	&	\tiny	 	0.015	 	&	\tiny	 	0	 	&	\tiny	 	0.02	\\ 
				\tiny	 4	 	&	\tiny	 	ASIC1	 	&	\tiny	 	0	 	&	\tiny	 	0	 	&	\tiny	 	0	 	&	\tiny	 	0	 	&	\tiny	 	0	 	&	\tiny	 	0.003	\\ 
				\tiny	 5	 	&	\tiny	 	BMPR2	 	&	\tiny	 	2	 	&	\tiny	 	0	 	&	\tiny	 	0.03	 	&	\tiny	 	0.104	 	&	\tiny	 	0	 	&	\tiny	 	0.02	\\ 
				\tiny	 6	 	&	\tiny	 	CFB	 	&	\tiny	 	2	 	&	\tiny	 	0	 	&	\tiny	 	0.03	 	&	\tiny	 	0.115	 	&	\tiny	 	0	 	&	\tiny	 	0.018	\\ 
				\tiny	 7	 	&	\tiny	 	CFH	 	&	\tiny	 	2	 	&	\tiny	 	0	 	&	\tiny	 	0.03	 	&	\tiny	 	0.121	 	&	\tiny	 	0	 	&	\tiny	 	0.018	\\ 
				\tiny	 8	 	&	\tiny	 	CHMP1B	 	&	\tiny	 	2	 	&	\tiny	 	0	 	&	\tiny	 	0.03	 	&	\tiny	 	0.094	 	&	\tiny	 	0	 	&	\tiny	 	0.021	\\ 
				\tiny	 9	 	&	\tiny	 	CLASP2	 	&	\tiny	 	1	 	&	\tiny	 	0	 	&	\tiny	 	0.015	 	&	\tiny	 	0.108	 	&	\tiny	 	0	 	&	\tiny	 	0.01	\\ 
				\tiny	 10	 	&	\tiny	 	CNNM2	 	&	\tiny	 	1	 	&	\tiny	 	0	 	&	\tiny	 	0.015	 	&	\tiny	 	0.077	 	&	\tiny	 	0	 	&	\tiny	 	0.012	\\ 
				\tiny	 11	 	&	\tiny	 	COPB1	 	&	\tiny	 	2	 	&	\tiny	 	0	 	&	\tiny	 	0.03	 	&	\tiny	 	0.121	 	&	\tiny	 	0	 	&	\tiny	 	0.019	\\ 
				\tiny	 12	 	&	\tiny	 	DDC	 	&	\tiny	 	0	 	&	\tiny	 	0	 	&	\tiny	 	0	 	&	\tiny	 	0	 	&	\tiny	 	0	 	&	\tiny	 	0.003	\\ 
				\tiny	 13	 	&	\tiny	 	EIF2B1	 	&	\tiny	 	0	 	&	\tiny	 	0	 	&	\tiny	 	0	 	&	\tiny	 	0	 	&	\tiny	 	0	 	&	\tiny	 	0.003	\\ 
				\tiny	 14	 	&	\tiny	 	EPHB4	 	&	\tiny	 	0	 	&	\tiny	 	0	 	&	\tiny	 	0	 	&	\tiny	 	0	 	&	\tiny	 	0	 	&	\tiny	 	0.003	\\ 
				\tiny	 15	 	&	\tiny	 	ERAP1	 	&	\tiny	 	0	 	&	\tiny	 	0	 	&	\tiny	 	0	 	&	\tiny	 	0	 	&	\tiny	 	0	 	&	\tiny	 	0.003	\\ 
				\tiny	 16	 	&	\tiny	 	ERBB2	 	&	\tiny	 	3	 	&	\tiny	 	0.333	 	&	\tiny	 	0.045	 	&	\tiny	 	0.137	 	&	\tiny	 	0.333	 	&	\tiny	 	0.023	\\ 
				\tiny	 17	 	&	\tiny	 	FABP6	 	&	\tiny	 	1	 	&	\tiny	 	0	 	&	\tiny	 	0.015	 	&	\tiny	 	0.114	 	&	\tiny	 	0	 	&	\tiny	 	0.01	\\ 
				\tiny	 18	 	&	\tiny	 	FAM20A	 	&	\tiny	 	0	 	&	\tiny	 	0	 	&	\tiny	 	0	 	&	\tiny	 	0	 	&	\tiny	 	0	 	&	\tiny	 	0.003	\\ 
				\tiny	 19	 	&	\tiny	 	GNPNAT1	 	&	\tiny	 	0	 	&	\tiny	 	0	 	&	\tiny	 	0	 	&	\tiny	 	0	 	&	\tiny	 	0	 	&	\tiny	 	0.003	\\ 
				\tiny	 20	 	&	\tiny	 	GP1BA	 	&	\tiny	 	0	 	&	\tiny	 	0	 	&	\tiny	 	0	 	&	\tiny	 	0	 	&	\tiny	 	0	 	&	\tiny	 	0.003	\\ 
				\tiny	 21	 	&	\tiny	 	GPC1	 	&	\tiny	 	1	 	&	\tiny	 	0	 	&	\tiny	 	0.015	 	&	\tiny	 	0.015	 	&	\tiny	 	0	 	&	\tiny	 	0.02	\\ 
				\tiny	 22	 	&	\tiny	 	GRB7	 	&	\tiny	 	2	 	&	\tiny	 	\textbf{1}	 	&	\tiny	 	0.03	 	&	\tiny	 	0.126	 	&	\tiny	 	1	 	&	\tiny	 	0.016	\\ 
				\tiny	 23	 	&	\tiny	 	GTF2B	 	&	\tiny	 	1	 	&	\tiny	 	0	 	&	\tiny	 	0.015	 	&	\tiny	 	0.108	 	&	\tiny	 	0	 	&	\tiny	 	0.01	\\ 
				\tiny	 24	 	&	\tiny	 	HEXA	 	&	\tiny	 	0	 	&	\tiny	 	0	 	&	\tiny	 	0	 	&	\tiny	 	0	 	&	\tiny	 	0	 	&	\tiny	 	0.003	\\ 
				\tiny	 25	 	&	\tiny	 	HLA-DRB1	 	&	\tiny	 	2	 	&	\tiny	 	0	 	&	\tiny	 	0.03	 	&	\tiny	 	0.132	 	&	\tiny	 	0	 	&	\tiny	 	0.017	\\ 
				\tiny	 26	 	&	\tiny	 	HSF1	 	&	\tiny	 	1	 	&	\tiny	 	0	 	&	\tiny	 	0.015	 	&	\tiny	 	0.124	 	&	\tiny	 	0	 	&	\tiny	 	0.01	\\ 
				\tiny	 27	 	&	\tiny	 	IHH	 	&	\tiny	 	1	 	&	\tiny	 	0	 	&	\tiny	 	0.015	 	&	\tiny	 	0.015	 	&	\tiny	 	0	 	&	\tiny	 	0.02	\\ 
				\tiny	 28	 	&	\tiny	 	IST1	 	&	\tiny	 	1	 	&	\tiny	 	0	 	&	\tiny	 	0.015	 	&	\tiny	 	0.081	 	&	\tiny	 	0	 	&	\tiny	 	0.012	\\ 
				\tiny	 29	 	&	\tiny	 	ITGB7	 	&	\tiny	 	1	 	&	\tiny	 	0	 	&	\tiny	 	0.015	 	&	\tiny	 	0.102	 	&	\tiny	 	0	 	&	\tiny	 	0.011	\\ 
				\tiny	 30	 	&	\tiny	 	KCNK1	 	&	\tiny	 	4	 	&	\tiny	 	0.333	 	&	\tiny	 	0.06	 	&	\tiny	 	0.163	 	&	\tiny	 	0.333	 	&	\tiny	 	0.028	\\ 
				\tiny	 31	 	&	\tiny	 	KCNK16	 	&	\tiny	 	3	 	&	\tiny	 	0.667	 	&	\tiny	 	0.045	 	&	\tiny	 	0.147	 	&	\tiny	 	0.667	 	&	\tiny	 	0.021	\\ 
				\tiny	 32	 	&	\tiny	 	KCNK17	 	&	\tiny	 	2	 	&	\tiny	 	\textbf{1}	 	&	\tiny	 	0.03	 	&	\tiny	 	0.129	 	&	\tiny	 	1	 	&	\tiny	 	0.015	\\ 
				\tiny	 33	 	&	\tiny	 	KCNQ1	 	&	\tiny	 	4	 	&	\tiny	 	0.167	 	&	\tiny	 	0.06	 	&	\tiny	 	0.171	 	&	\tiny	 	0.167	 	&	\tiny	 	0.027	\\ 
				\tiny	 34	 	&	\tiny	 	KPNA2	 	&	\tiny	 	0	 	&	\tiny	 	0	 	&	\tiny	 	0	 	&	\tiny	 	0	 	&	\tiny	 	0	 	&	\tiny	 	0.003	\\ 
				\tiny	 35	 	&	\tiny	 	MAPT	 	&	\tiny	 	7	 	&	\tiny	 	0.048	 	&	\tiny	 	0.104	 	&	\tiny	 	0.132	 	&	\tiny	 	0.048	 	&	\tiny	 	0.054	\\ 
				\tiny	 36	 	&	\tiny	 	MX1	 	&	\tiny	 	1	 	&	\tiny	 	0	 	&	\tiny	 	0.015	 	&	\tiny	 	0.124	 	&	\tiny	 	0	 	&	\tiny	 	0.01	\\ 
				\tiny	 37	 	&	\tiny	 	NDUFB5	 	&	\tiny	 	2	 	&	\tiny	 	\textbf{1}	 	&	\tiny	 	0.03	 	&	\tiny	 	0.109	 	&	\tiny	 	1	 	&	\tiny	 	0.017	\\ 
				\tiny	 38	 	&	\tiny	 	NDUFV1	 	&	\tiny	 	2	 	&	\tiny	 	\textbf{1}	 	&	\tiny	 	0.03	 	&	\tiny	 	0.109	 	&	\tiny	 	1	 	&	\tiny	 	0.017	\\ 
				\tiny	 39	 	&	\tiny	 	NIPA1	 	&	\tiny	 	2	 	&	\tiny	 	0	 	&	\tiny	 	0.03	 	&	\tiny	 	0.089	 	&	\tiny	 	0	 	&	\tiny	 	0.022	\\ 
				\tiny	 40	 	&	\tiny	 	NIPAL1	 	&	\tiny	 	0	 	&	\tiny	 	0	 	&	\tiny	 	0	 	&	\tiny	 	0	 	&	\tiny	 	0	 	&	\tiny	 	0.003	\\ 
				\tiny	 41	 	&	\tiny	 	NIPAL4	 	&	\tiny	 	0	 	&	\tiny	 	0	 	&	\tiny	 	0	 	&	\tiny	 	0	 	&	\tiny	 	0	 	&	\tiny	 	0.003	\\ 
				\tiny	 42	 	&	\tiny	 	NR1H4	 	&	\tiny	 	5	 	&	\tiny	 	0	 	&	\tiny	 	0.075	 	&	\tiny	 	0.14	 	&	\tiny	 	0	 	&	\tiny	 	0.044	\\ 
				\tiny	 43	 	&	\tiny	 	OBP2B	 	&	\tiny	 	0	 	&	\tiny	 	0	 	&	\tiny	 	0	 	&	\tiny	 	0	 	&	\tiny	 	0	 	&	\tiny	 	0.003	\\ 
				\tiny	 44	 	&	\tiny	 	PDPK1	 	&	\tiny	 	2	 	&	\tiny	 	0	 	&	\tiny	 	0.03	 	&	\tiny	 	0.093	 	&	\tiny	 	0	 	&	\tiny	 	0.017	\\ 
				\tiny	 45	 	&	\tiny	 	PHF1	 	&	\tiny	 	2	 	&	\tiny	 	0	 	&	\tiny	 	0.03	 	&	\tiny	 	0.11	 	&	\tiny	 	0	 	&	\tiny	 	0.018	\\ 
				\tiny	 46	 	&	\tiny	 	PPP3CA	 	&	\tiny	 	0	 	&	\tiny	 	0	 	&	\tiny	 	0	 	&	\tiny	 	0	 	&	\tiny	 	0	 	&	\tiny	 	0.003	\\ 
				\tiny	 47	 	&	\tiny	 	PSMC2	 	&	\tiny	 	5	 	&	\tiny	 	0.1	 	&	\tiny	 	0.075	 	&	\tiny	 	0.149	 	&	\tiny	 	0.1	 	&	\tiny	 	0.039	\\ 
				\tiny	 48	 	&	\tiny	 	PSMD13	 	&	\tiny	 	1	 	&	\tiny	 	0	 	&	\tiny	 	0.015	 	&	\tiny	 	0.119	 	&	\tiny	 	0	 	&	\tiny	 	0.01	\\ 
				\tiny	 49	 	&	\tiny	 	PTPN11	 	&	\tiny	 	4	 	&	\tiny	 	0.167	 	&	\tiny	 	0.06	 	&	\tiny	 	0.149	 	&	\tiny	 	0.167	 	&	\tiny	 	0.029	\\ 
				\tiny	 50	 	&	\tiny	 	RAB2A	 	&	\tiny	 	1	 	&	\tiny	 	0	 	&	\tiny	 	0.015	 	&	\tiny	 	0.1	 	&	\tiny	 	0	 	&	\tiny	 	0.011	\\ 
				\tiny	 51	 	&	\tiny	 	RARA	 	&	\tiny	 	3	 	&	\tiny	 	0	 	&	\tiny	 	0.045	 	&	\tiny	 	0.132	 	&	\tiny	 	0	 	&	\tiny	 	0.025	\\ 
				\tiny	 52	 	&	\tiny	 	RBM6	 	&	\tiny	 	0	 	&	\tiny	 	0	 	&	\tiny	 	0	 	&	\tiny	 	0	 	&	\tiny	 	0	 	&	\tiny	 	0.003	\\ 
				\tiny	 53	 	&	\tiny	 	RPS26	 	&	\tiny	 	3	 	&	\tiny	 	0.333	 	&	\tiny	 	0.045	 	&	\tiny	 	0.161	 	&	\tiny	 	0.333	 	&	\tiny	 	0.022	\\ 
				\tiny	 54	 	&	\tiny	 	RPS6KA3	 	&	\tiny	 	2	 	&	\tiny	 	0	 	&	\tiny	 	0.03	 	&	\tiny	 	0.109	 	&	\tiny	 	0	 	&	\tiny	 	0.017	\\ 
				\tiny	 55	 	&	\tiny	 	RPSA	 	&	\tiny	 	2	 	&	\tiny	 	\textbf{1}	 	&	\tiny	 	0.03	 	&	\tiny	 	0.138	 	&	\tiny	 	1	 	&	\tiny	 	0.016	\\ 
				\tiny 56	&\tiny	SLC27A5	&\tiny	1	&\tiny	0	&\tiny	0.015	&\tiny	0.114	&\tiny	0	&\tiny	0.01	\\
				\tiny 57	&\tiny	SLC4A4	&\tiny	0	&\tiny	0	&\tiny	0	&\tiny	0	&\tiny	0	&\tiny	0.003	\\
				\tiny 58	&\tiny	SNRPB	&\tiny	2	&\tiny	\textbf{1}	&\tiny	0.03	&\tiny	0.03	&\tiny	1	&\tiny	0.02	\\
				\tiny 59	&\tiny	SNRPD1	&\tiny	2	&\tiny	\textbf{1}	&\tiny	0.03	&\tiny	0.03	&\tiny	1	&\tiny	0.02	\\
				\tiny 60	&\tiny	SNRPE	&\tiny	2	&\tiny	\textbf{1}	&\tiny	0.03	&\tiny	0.03	&\tiny	1	&\tiny	0.02	\\
				\tiny 61	&\tiny	TERF2IP	&\tiny	0	&\tiny	0	&\tiny	0	&\tiny	0	&\tiny	0	&\tiny	0.003	\\
				\tiny 62	&\tiny	TNFRSF13C	&\tiny	0	&\tiny	0	&\tiny	0	&\tiny	0	&\tiny	0	&\tiny	0.003	\\
				\tiny 63	&\tiny	TRPM1	&\tiny	0	&\tiny	0	&\tiny	0	&\tiny	0	&\tiny	0	&\tiny	0.003	\\
				\tiny 64	&\tiny	UBE2I	&\tiny	5	&\tiny	0	&\tiny	0.075	&\tiny	0.157	&\tiny	0	&\tiny	0.04	\\
				\tiny 65	&\tiny	UGT1A3	&\tiny	1	&\tiny	0	&\tiny	0.015	&\tiny	0.114	&\tiny	0	&\tiny	0.01	\\
				\tiny 66	&\tiny	UGT2B7	&\tiny	0	&\tiny	0	&\tiny	0	&\tiny	0	&\tiny	0	&\tiny	0.003	\\
				\tiny 67	&\tiny	YWHAH	&\tiny	2	&\tiny	0	&\tiny	0.03	&\tiny	0.109	&\tiny	0	&\tiny	0.017	\\
				\tiny 68	&\tiny	ZRANB3	&\tiny	0	&\tiny	0	&\tiny	0	&\tiny	0	&\tiny	0	&\tiny	0.003	\\
				
				\botrule
			\end{tabular}
		\footnotetext[1]{ \tiny Significant proteins:- AHSG: Alpha 2-HS Glycoprotein,
			KCNK1: Potassium channel subfamily K member 1,
			KCNQ1: Potassium Voltage-Gated Channel Subfamily Q Member 1,
			MAPT: Microtubule Associated Protein Tau,
			NR1H4: Nuclear Receptor Subfamily 1 Group H Member 4,
			PSMC2: Proteasome 26S Subunit, ATPase 2,
			PTPN11: Tyrosine-protein phosphatase non-receptor type 11,
			UBE21: UBE2I ubiquitin conjugating enzyme E2 I}
			
		\end{minipage}
	\end{center}
\end{table}

\clearpage

\section{Conclusion}\label{sec13}
Centrality analysis are very useful for analyzing large biological networks. Using a candidate gene network of Omicron as a case study, we investigated and compared different centrality measures. According to the findings, it is beneficial to explore candidate gene networks using methods from other fields of science such as social network analysis. On the $7$ base lineage of Omicron variations, including the $68$ unique protein encoded by the Omicron candidate gene, graph analysis is done. From the Omicron main network, we extracted the Markov clustering algorithm's findings i.e., $18$ network clusters. The primary Omicron network has $68$ nodes, each of which represents a protein. Of the $68$ proteins, $8$ were found to be significant, including AHSG, KCNK1, KCNQ1, MAPT, NR1H4, PSMC2, PTPN11, and UBE21, with the MAPT protein receiving the highest score. The MAPT protein has the most dominating influence on the protein-protein interaction network of the Omicron candidate gene, according to the centrality score.
Medical researchers as well as the general public will benefit from this work, as it will allow them to  to consider biological knowledge in network analysis of the Omicron virus.

Analysis of networks can benefit greatly from centrality measures. However, they are also required to be properly informed, selected, and applied. As part of our main research work, we present information about the four major centrality measures that have been found to be relevant for finding the most significant proteins in the Omicron Lineage Variants PPIN. A wide range of new and large networks are being created and developed due to different applications and different centrality measures. The majority of studies have tried to demonstrate the uniqueness and superiority of their centrality measures. We still have a lot to learn about making a difference and applying them properly. This is how we presented it.

\begin{itemize}
	\item  \textbf{Authors and Affiliations:} 
	\begin{itemize}
		\item Mamata Das, Department of Computer Applications, NIT Trichy, India 
		\item Selvakumar K., Department of Computer Applications, NIT Trichy, India 
		\item P.J.A. Alphonse, Department of Computer Applications, NIT Trichy, India 
		\item \textbf{Corresponding author:} Mamata Das and Selvakumar K.
	\end{itemize}
  

\end{itemize}
\vspace{-2mm}
\vspace{-1mm}
\begin{itemize}
	\item  \textbf{Compliance with Ethical Standards} 
	\begin{itemize}
		\item \textbf{Funding:} This is the work of the first author under her doctoral. This research received no external funding.
		\item \textbf{Disclosure of potential conflicts of interest:} On behalf of all authors, the corresponding author states that there is no conflict of interest.
		\item \textbf{Research involving human participants and/or animals:} This article does not contain any studies with human participants or animals performed by any of the authors.
	\end{itemize}

\end{itemize}

{\small
\bibliographystyle{plain}
\bibliography{24082022_PPIcentrality}
}


\end{document}